\documentclass[12pt]{article}
\usepackage{amssymb}
\usepackage[mathscr]{eucal}
\usepackage{url}
\usepackage{a4}
\usepackage{cite}
\usepackage{graphicx}
\usepackage{psfrag}
\usepackage{subfigure}
\usepackage{latexsym}
\usepackage{amsmath}

\newcounter{mnotecount}[section]

\begin{document}
\newcommand{\g}{$\bf \bar g g^{\alpha\beta}$}
\title{Mechanics of floating bodies
}
\author{Robert Beig\\Gravitational Physics\\Faculty of Physics, University
of Vienna\\ Boltzmanngasse 5, A-1090 Vienna, Austria\\[1cm]
Bernd G. Schmidt\\
Max-Planck-Institut f\"ur Gravitationsphysik\\ Albert-Einstein-Institut\\
Am M\"uhlenberg 1, D-14476 Golm, Germany}

\maketitle

\begin{abstract}
We introduce and study the mechanical system which describes the dynamics and statics of rigid bodies of constant density floating in a calm incompressible fluid. Since much of the standard equilibrium theory, starting with Archimedes, allows bodies with vertices and edges, we assume the bodies to be convex and take care not to assume more regularity than that implied by convexity. One main result is the (Liapunoff) stability of equilibria satisfying a condition equivalent to the standard 'metacentric' criterion.

\end{abstract}

\section{Introduction}
Equilibria of floating bodies have been studied since antiquity (see \cite{A}, \cite{Ro} and references therein). Results on their stability and the associated concept of the metacentre date back to the 18th century  (see \cite{F} for a modern presentation) and continue to play a role in areas such as naval architecture \cite{T} or the study of icebergs \cite {P}. This field has recently also regained mathematical interest. In part this is in pursuit of the  famous question of Ulam \cite{U} if there are bodies other than spheres which have equilibria in every orientation \cite{G,V,R,R1}. In part classifying the equilibria of different bodies such as  certain polyhedra \cite{G,Erd1} and their stability poses mathematical challenges. In fact, even the treatment of the floating paraboloid originally studied by Archimedes himself still leaves room for clarifications \cite{GK}.\\ \\
There is on the other hand, as has also been pointed out in \cite{MV},  an associated time dependent theory described by a conservative mechanical system - and that, to the best of our knowledge, has received no rigorous study whatsoever. The present paper attempts to be at least a starting point for filling this gap in the literature. \\ \\
The dynamical system of the floating body has some analogy with, but more structure than, the heavy top\footnote{Recall the heavy top is simply a top subject to a constant gravitational field with one point fixed, see e.g. \cite{MR}.} - in three respects. First there is for the floating body an additional physical parameter, namely the density of the body relative to that of water. Second, there is a translational degree of freedom not present for the heavy top afforded by the height of the floating body. Third, the torque due to gravity acting on the heavy top, in the comoving frame, is essentially the vector product of the instantaneous direction of gravity with the given vector which points from the fixed point of the body to its center of mass. For the floating body, by Archimedes' principle, the role of that latter vector is played by the vector, in the comoving frame, connecting the center of mass with the buoyancy center, i.e. the centroid of the submerged part of the body - and that is dynamical and depends on the detailed shape of the body. So, for example, the floating body can have a wealth of equilibrium configurations, whereas the heavy top has just two, namely the unstable upright and the stable hanging one\footnote{By equilibria we always mean solutions which are strictly static - not the 'relative equilibria' associated with steady rotations.}.\\ \\
We start, in Sect.2, with a crash-course on the rigid body in an external force field. One motivation for doing this in spite of the vast literature on the subject is to fix notation and to carefully distinguish between the 'material space' and 'physical space': this is customary in the field of continuum mechanics, and it avoids confusions which treatments in the standard literature on rigid body mechanics can lead to - and especially so in the present context. The book closest to our approach is \cite{MR}. We also derive the equations for the heavy top, which is useful for comparisons with the floating body.\\ \\
In Sect.3 we specialize to the force field according to the principle of Archimedes: a constant field of strength $g$ pointing  along the downward normal to the water plane plus a constant, upward-pointing buoyancy field supported under the water plane of strength $g/s$, where $s$ is the density of the body relative to that of water.
Let us remark at this point that in so doing we  are making many simplifying assumptions: We discard hydrodynamic effects such as the presence of viscosity or vorticity, related to phenomena such as dynamical buoyancy ('lift') acting on the body. We are not taking into account the work done by an accelerating body against the inertia of the fluid as first described by Kirchhoff \cite{K} for a fluid filling all of space. And of course we ignore the backreaction on the body due to surface waves generated by the motion of the body as first treated in \cite{J}.\\
Applying the scheme of Sect.2 and factoring out horizontal translational motions, which decouple from the other degrees of freedom, results in a dynamical system in the 7 variables $(h,\dot{h}, \vec{\Pi}, \vec{n}) \in \mathbb{R}^1 \times \mathbb{R}^1 \times \mathbb{R}^3 \times \mathbb{S}^2$, where $h$ is the height of the center of mass relative to the water plane, $\vec{\Pi}$ is the body-angular momentum and the unit-vector $\vec{n}$ is the direction of gravity in the body frame. There is a conserved energy with  kinetic term
$\frac{\dot{h}^2}{2} + \frac{(I^{- 1} \vec{\Pi},\vec{\Pi})}{2}$, where $I$ the tensor of inertia, plus a potential term $\mathcal{V}_s (h, \vec{n})$, which describes the deviation of the motion from that of a free top. The  dependence of the potential comes from the fact that the full Euclidean symmetry of the free rigid body under  $\mathbb{E}(3)$ is due to the presence of gravity and buoyancy broken down to $\mathbb{E}(2)$ and $\{(h, \vec{n})\} =
\mathbb{R}^1 \times \mathbb{S}^2$ is equal to the factor space
$\mathbb{E}(3)/\mathbb{E}(2)$. In the case of the heavy top, where the potential just depends on $\vec{n}$, the relationship between the Hamiltonian and the equations of motion is via a Poisson structure $\{,\}$ on $\mathbb{R}^3 \times \mathbb{S}^2$ \cite{MR}. When our equations are written in first-order form and after the 2 'symplectic' dimensions due to the degree of freedom afforded by $h$ have been added to the Poisson structure of the heavy top, it turns out that our equations of motion are also of Poisson type w.r. to the given Hamiltonian, but that viewpoint is not further pursued here.\\

Our work in the subsequent sections centers around and is based on the potential $\mathcal{V}_s(h,\vec{n})$, in particular its regularity properties for a compact, convex body. For a given body $\mathbb{R}^1 \times \mathbb{S}^2$ splits into three regions: two corresponding to the body being up in the air or completely immersed. The third, intermediate, region is where the body is partially immersed, and that is the interesting one, because only here there is an interaction between the vertical ($h$-) degree of freedom and the directional ($\vec{n}$-) degree of freedom: in the first-named regions there is just a constant force field which exerts no torque on a rigid body in its rest frame. And, of course, for $s < 1$, critical points of the potential - which correspond to equilibrium configurations - all occur in the intermediate region. It now turns out, that $\mathcal{V}_s(h,\vec{n})$ is $C^2$ in this region even for a general convex body: this justifies the use of the Hessian of the potential, whose definiteness properties characterize stability. And it also justifies the use of the Poincar\'e-Hopf theorem first advocated in the elegant paper \cite{G} - in particular, as done there, for bodies with vertices and edges. It also turns out that the potential is smooth in the partially immersed region for smooth bodies, but globally at best $C^2$ across its boundary, whose points correspond to the body touching the water plane from above or below. On the other hand the potential is $C^{1,1}$, i.e. has H\"older continuous first derivatives, globally for all convex bodies, which implies that solutions to the dynamical equations exist and are unique.\\ \\
In Sect.5 we treat the equilibrium theory, work out the Hessian of $\mathcal{V}_s$ at a critical point and prove the Liapunoff stability of solutions evolving from stable critical points. Finally, in Sect.6, we calculate  $\mathcal{V}_s$ explicitly in the case of the body being a triaxial ellipsoid and point out its most elementary consequences such as its equilibria (confirming results in \cite{G}) and the simplest time dependent solutions. Clearly our findings in this section can only be a starting point for - and hopefully an invitation for others to undertake - efforts to reach a full understanding of the allowed solutions of the dynamical equations for the ellipsoid or more general shapes.
In Sect. 7 we summarize our findings.
The Appendix supplies the basic argument for the regularity of the potential.

\section{The rigid body}
For the sake of conceptual clarity we start with a quick review of the rigid body. We take rigid body motions to be time-dependent, orientation-preserving, one-one  maps $\phi : \mathcal{B} \in (\mathbb{R}^3_\mathcal{B}, \delta_{AB}) \rightarrow (\mathbb{R}^3_\mathcal{S},\delta_{ij})$, where $\mathbb{R}^3_\mathcal{B}$, resp. $\mathbb{R}^3_\mathcal{S}$ are the material, resp. spatial copies of flat $\mathbb{R}^3$ and $\mathcal{B}$ a compact, connected domain in the former. We are throughout using the Einstein summation convention and indices $A,B,..$ for quantities on material and $i,j,..$ for quantities on physical space. Capital indices are lowered and raised with $\delta_{AB}$, lowercase indices with $\delta_{ij}$. The volume elements on both spaces are respectively written as $\epsilon_{ABC}$ and $\epsilon_{ijk}$. We will assume that the centroid ('center of gravity') of $\mathcal{B}$ is the origin, i.e.
\begin{equation}\label{centroid}
\int_\mathcal{B} X^A d^3X = 0\,.
\end{equation}
The maps $\phi$ are subject to the constraint that they be isometries between these spaces. Thus (see \cite{MU}) in Euclidean coordinates on both spaces they take the form $x^i = \phi^i(t,X) = c^i(t) + R^i{}_A (t) X^A$, where
\begin{equation}\label{constraint}
R^i{}_A (t) R^j{}_B (t)\,\delta_{ij} = \delta_{AB}\,\,\,\Leftrightarrow\,\,\,
R^i{}_A (t) R^j{}_B (t)\,\delta^{AB} = \delta^{ij}
\end{equation}
Next define the material angular velocity $\Omega^A$ by
\begin{equation}\label{time}
\delta_{ij} \,R^i{}_A \dot{R}^j{}_B = - \epsilon_{ABC}\Omega^C\,\,\,\Leftrightarrow\,\,\,\dot{R}^i{}_A + \epsilon_{AB}{}^C \Omega^B R^i{}_C = 0
\end{equation}
(We will here and throughout use a dot or $\frac{d}{dt}$ for time derivative interchangibly depending on typographical convenience.) Together with (\ref{constraint}) it follows that
\begin{equation}
\delta_{ij}\,\dot{R^i}{}_A \dot{R}^j{}_B  = (\delta_{AB}\delta_{CD} - \delta_{AC}\delta_{BD})\, \Omega^C \Omega^D\,.
\end{equation}
We assume that $R^i{}_A$ also preserves orientation, i.e. that
\begin{equation}\label{orient}
R^i{}_A R^j{}_B R^k{}_C \,\epsilon_{ijk} = \epsilon_{ABC}\,,
\end{equation}
which implies
\begin{equation}\label{epsilonRR}
\epsilon^i{}_{jk}R^j{}_A \dot{R}^k{}_B = 2 \,\delta_{A[B} R^i{}_{C]} \,\Omega^C\,.
\end{equation}
We henceforth restrict ourselves to bodies of constant density. The latter condition will be no restriction for the force we are interested in.
Now the (linear) momentum $P^i$ of the body arises by evaluating
the momentum of a point particle, namely the quantity $\dot{x}^i$ along $\phi^i$ and integrating the result  over $\mathcal{B}$. Thus
\begin{equation}
|\mathcal{B}|\, P^i = \int_\mathcal{B}(\dot{c}^i + \dot{R}^i{}_A X^a) d^3X =
|\mathcal{B}|\,\dot{c}^i\,,
\end{equation}
where we have used (\ref{centroid}). For the angular momentum $L^i$ we obtain a sum of orbital and intrinsic angular momentum. Namely, using (\ref{epsilonRR}) for the latter, we obtain the expression
\begin{equation}
 L^i = \epsilon^i{}_{jk} c^j \dot{c}^k + R^i{}_A I^A{}_B \Omega^B\,,
\end{equation}
where $I_{AB} = |\mathcal{B}|^{- 1} \int_\mathcal{B} [\delta_{AB}(X,X) - X_A X_B]\, d^3X$ is the inertia tensor of $\mathcal{B}$. The equations of motion for a rigid body with constant density and  mass 1 in a force field $F^i(x)$ then take the form
\begin{equation}
\frac{d P}{dt}^i = \mathcal{F}^i\,,\hspace{1cm}\frac{d L}{dt}^i = \mathcal{M}^i\,,
\end{equation}
where $\mathcal{F}^i$, the total force, is given by $\mathcal{F}^i (c,R) = |\mathcal{B}|^{- 1}\int_\mathcal{B} F^i(c + RX) d^3X$ and the total torque
is given by $\mathcal{M}^i (c,R) = |\mathcal{B}|^{- 1}\int_\mathcal{B} \epsilon^i{}_{j}{}^k(c^j + R^j{}_A X^A)F_k(c + RX) d^3X$.
Thus
\begin{equation}\label{ddotc}
\frac{d^2 c^i}{dt^2} = |\mathcal{B}|^{- 1}\int_\mathcal{B} F^i(c + RX)\, d^3X
\end{equation}
and, using  (\ref{ddotc}), the $\ddot{c}$-term in the torque equation drops out, and we obtain
\begin{equation}\label{torque}
(R^i{}_A I^A{}_B \Omega^B)^{.} = |\mathcal{B}|^{- 1}\int_\mathcal{B} \epsilon^i{}_{j}{}^k R^j{}_A X^A F_k(c + RX) d^3X\,.
\end{equation}
Now, using (\ref{time}), the l.h. side of (\ref{torque}) can be written as
\begin{equation}
R^i{}_A (I^A{}_B \dot{\Omega}^B + \epsilon^A{}_{BC} \Omega^B I^C{}_D \Omega^D)
\end{equation}
and, using (\ref{orient}) the r.h. side can be written as
\begin{equation}
 |\mathcal{B}|^{- 1}\int_\mathcal{B} R^i{}_A \,\epsilon^A{}_B{}^C X ^B  R^j{}_C F_j(c + RX) d^3X\,.
\end{equation}
Thus, introducing the notation $\vec{\Pi} = I \vec{\Omega}$ and $(\vec{f}\,)_A = R^i{}_A F_i$, the torque equation can be written as
\begin{equation}\label{torque1}
\frac{d \vec{\Pi}}{dt}+ \vec{\Omega}  \times  \vec{\Pi} = |\mathcal{B}|^{- 1}\int_\mathcal{B} \vec{X} \times \vec{f}\,(c + RX) \,d^3X\,.
\end{equation}
Adding (\ref{time}), namely
\begin{equation}\label{time1}
\dot{R}^i{}_A + \epsilon_{AB}{}^C \Omega^B R^i{}_C = 0\,,
\end{equation}
the equations (\ref{ddotc}, \ref{torque1}, \ref{time1}) form a closed system for the variables $(c^i, \Pi^A, R^i{}_A)$. There is no need to introduce coordinates on $\mathbb{SO}(3)$: the condition that the linear map $R^i{}_A$ be in $\mathbb{SO}(3)$ can be merely imposed as initial condition. Alternatively, if one does use a parametrization of $\mathbb{SO}(3)$ such as the Euler angles, the 6 relations (\ref{time1}) boil down to the 3 identities which express $\Omega^A$ in terms the Euler angles and their first time derivatives (see e.g.\cite{MR}).\\ \\
We add two remarks.\\ \\
Remark 1: If the force field has a potential, i.e. $F_i(x) = - \partial_i U(x)$, the equations (\ref{ddotc},\ref{torque1}) can be obtained by first evaluating, as done previously with momentum and angular momentum, the Lagrangian
\begin{equation}
L(x,\dot{x}) = \frac{(\dot{x},\dot{x})}{2} - U(x)
\end{equation}
along the maps $x^i =\phi^i(t,X) = c^i(t) + R^i{}_A(t) X^A$
resulting in $|\mathcal{B}|$ times the expression
\begin{equation}
\mathcal{L}(c,\dot{c},R, \dot{R}) = \frac{1}{2} \delta_{ij} \dot{c}^i \dot{c}^j + \frac{1}{2} I_{AB} \,\Omega^A \Omega^B - |\mathcal{B}|^{- 1} \int_\mathcal{B} U (c + RX) \,d^3X\,.
\end{equation}
One then varies the action $\int \mathcal{L} dt$ w.r. to $c^i$ and w.r. to $R^i{}_A$, the latter variation under the constraint (\ref{constraint}), e.g. using Lagrange parameters. One also finds there is a conserved energy of the form
\begin{equation}
\mathcal{H}(c^i, \dot{c}^i, R^i{}_A,\vec{\Pi}) = \frac{(\dot{c},\dot{c})
+ (I^{- 1}\vec{\Pi}, \vec{\Pi})}{2} + |\mathcal{B}|^{- 1} \int_\mathcal{B} U (c + RX) \,d^3X\,.
\end{equation}
Finally, when the potential has a translational symmetry, i.e. $d^i \partial_i U = 0$ for some constant vector $d$, the associated component of
linear momentum, i.e. $(d, \dot{c})$ is conserved. When $U$ has a rotational symmetry, i.e. for some constant vector $m$ there holds $m^i \epsilon_{ij}{}^k x^j \partial_k U = 0$, the associated component of the total angular momentum, i.e. $m_i (\epsilon^i{}_{jk} c^j \dot{c}^k + R^i{}_A \Pi^A)$ is conserved. \\ \\
Remark 2: The case of the heavy top in the present setting is obtained as follows: consider as potential the function $U(x) = g (x, e_3)$ with $g$ a positive constant and require as additional constraint on the maps $\phi$ that they map a fixed vector $\xi^A$ in $\mathbb{R}_{\mathcal{B}}^3$ to the origin in $\mathbb{R}_{\mathcal{S}}^3$, i.e. $\phi^i(t,X) = R^i{}_A (t) (X^A - \xi^A)$. The conserved energy then becomes
\begin{equation}\label{convention}
\mathcal{H}(\vec{n},\vec{\Pi}) = \frac{1}{2} (\bar{I}^{- 1} \vec{\Pi},\vec{\Pi}) + g\,(\vec{n}, \vec{\chi})\,
\end{equation}
where $n_A = R^i{}_A (e_3)_i$, $\bar{I}_{AB}$ is the inertia tensor w.r. to the fixed point and $\chi^A = - \xi^A$ is the vector pointing from
the fixed point to the centre of gravity. The equations of motion are
\begin{equation}\label{heavy}
\frac{d \vec{\Pi}}{dt}+ \vec{\Omega}  \times  \vec{\Pi} = \frac{g}{|\mathcal{B}|} \,\vec{n} \times \vec{\chi}
\end{equation}
Note that $R^i{}_A$ appears in (\ref{heavy}) only through $n_A$. Thus, in order to obtain a closed system, we can replace (\ref{time1}) by the simpler equation
\begin{equation}\label{n}
\frac{d \vec{n}}{d t} + \vec{\Omega} \times \vec{n} = 0\,.
\end{equation}
In addition to $\mathcal{H}$ the spin w.r. to the $e_3$-axis, i.e. $(\vec{n},\vec{\Pi})$, is conserved, as one easily checks. A simplification occurs in the presence of 'material axial symmetry'. Namely, suppose that $\vec{\chi}$ is an inertia axis and the remaining moments of inertia are equal: it is then not hard to see that these conditions are equivalent to the single equation
\begin{equation}
\chi^C \epsilon_{CD(A}\bar{I}^D{}_{B)} = 0
\end{equation}
and that there is then the additional conserved quantity $(\vec{\chi},\vec{\Pi})$. This is the case of the Lagrangian top, which is known to be integrable.
\section{The floating body}
We now specialize to the force field of our interest: this is a linear combination of a downward, homogenous gravitational field and an upward buoyancy force. We take, following Archimedes,
\begin{equation}
F^i(x) = - g \,(e_3)^i [1 - \frac{1}{s}\Theta(- (e_3,x))]\,,
\end{equation}
where $\Theta : \mathbb{R}^1 \rightarrow \mathbb{R}^1$ is the step function defined by $\Theta(u) = 1$ for $u \geq 0$ and $\Theta (u) = 0$ for $u < 0$ and $e_3$ is the unit vector pointing into the positive $x^3$-direction. Furthermore $g > 0$ is the constant gravitational acceleration and $s > 0$ the ratio of constant mass densities $s = \rho_{body}/\rho_{fluid}$. The water fills the half plane $(e_3,x) < 0$. This force has a potential, namely
\begin{equation}
U_s(x) = g (e_3,x)[1 - \frac{1}{s} \Theta(- (e_3,x))]\,.
\end{equation}
Since the potential is constant in $(x^1,x^2)$-planes, translational motion is inertial and will be ignored. We thus take
\begin{equation}
x^i = \phi^i(t,X) = h(t)(e_3)^i + R^i{}_A(t) X^A\,.
\end{equation}
Going through the steps described in the previous section, we obtain the reduced Hamiltonian
\begin{equation}\label{dynamics}
\mathcal{H} = \frac{\dot{h}^2}{2}+ \frac{(I^{- 1}\vec{\Pi}, \vec{\Pi})}{2} + \mathcal{V}_s(h,\vec{n})\,,
\end{equation}
where $\mathcal{V}_s (h,\vec{n}) = \int_\mathcal{B} U(\phi(t,X))\, d^3X$ is given by
\begin{equation}\label{potential1}
\mathcal{V}_s (h,\vec{n}) = g h\left(\!1 - \frac{|\mathcal{B}_{h,\vec{n}}|}{s |\mathcal{B}|}\right) + \frac{g}{s |\mathcal{B}|} \int_{\mathcal{B}_{h,\vec{n}}}\!\!\!\!\!\!(\vec{n},\vec{X})  d^3\!X = g \!\! \,\left[h - \frac{1}{s |\mathcal{B}| }\int_{\mathcal{B}_{h,\vec{n}}}\!\!\!\!\!\! (h - (\vec{n},\vec{X}))\,  d^3\!X \right]
\end{equation}
and where $\mathcal{B}_{h,\vec{n}}$ is defined as
\begin{equation}
\mathcal{B}_{h,\vec{n}} = \mathcal{B} \cap \{\vec{X}\,|\,h - (\vec{n},\vec{X}) \leq 0\}
\end{equation}
and $n_A$ is the normal to the water plane, as seen from the body and pointing into the water:
\begin{equation}
n_A = - R^i{}_A (e_3)_i\,,
\end{equation}
where we note the different convention from that entering (\ref{convention}).
The set $\mathcal{B}_{h,\vec{n}}$ can be described as the submerged part of the body, when the water plane as seen by the body has height $h$ relative the centroid of the body, where $h$ positive means that the centroid is above the water plane. In (\ref{potential1}) we have deliberatively written  $\mathcal{V}_s (h,\vec{n})$ in two forms. The reason is that the first form displays the positivity properties of the potential by splitting it into terms which are only $C^1$ separately, but in the second equation the two terms are $C^2$: all of this is explained in the next section.\\ \\
The equations of motion, from (\ref{ddotc}) and (\ref{torque1}), are
\begin{equation}\label{h}
\frac{d^2 h}{d t^2} = - g\,\left(1 - \frac{|\mathcal{B}_{h,\vec{n}}|}{s |\mathcal{B}|}\right)
\end{equation}
and
\begin{equation}\label{Pi}
\frac{d \vec{\Pi}}{dt}+ \vec{\Omega}  \times  \vec{\Pi} =  \frac{g}{s |\mathcal{B}|}\,\,\vec{n}\times\!\int_{\mathcal{B}_{h,\vec{n}}}\!\!\!\!\!\!\! \vec{X}\, d^3X
\end{equation}

As in the case of the heavy top, in order to obtain a closed system, we can replace (\ref{time1}) by the simpler equation
\begin{equation}\label{n3}
\frac{d \vec{n}}{d t} + \vec{\Omega} \times \vec{n} = 0\,.
\end{equation}
And as  in the case of the heavy top we have the additional conserved quantity $(\vec{n},\vec{\Pi})$. Again, there is a choice how to interpret the quantity $n^A$ in this system: we can either take $n^A$ as an element of $\mathbb{R}^3 \setminus \{0\}$ and pose $(\vec{n},\vec{n}) = 1$ as initial condition, which is propagated by virtue of (\ref{n}). Or we view $n^A$ as an element in $\mathbb{S}^2$: then equation (\ref{n3}) boils down to 2 evolution equations, say for spherical coordinates $(\theta,\phi)$.\\ \\
As opposed to the case of the heavy top, there does not seem to be the possibility for an additional conserved quantity when $\mathcal{B}$ is axially symmetric except in the relatively trivial case, where $\mathcal{B}$ is a sphere  and in which case the equations (\ref{ddotc}) and (\ref{torque1}) decouple and  (\ref{torque1}) becomes the equation of a free spherical top.

\section{Properties of $\mathcal{V}_s(h,\vec{n})$ and implications}

From now on we suppose in addition that the body $\mathcal{B}$ is convex. There are then functions \cite {HW} $h_-(\vec{n}) < 0 < h_+(\vec{n})$ so that the set $\mathcal{B}_{h,\vec{n}}$ touches the water plane from below at $h = h_-$ and from above at $h = h_+$. When $h > h_+$, the body is 'up in the air', i.e.  $\mathcal{B}_{h,\vec{n}}$ is empty. When $h_- < h < h_+$, the body is partially immersed and when $h < h_-$ the body is completely immersed, i.e. $\mathcal{B}_{h,\vec{n}} = \mathcal{B}$. Analytically $h_-$ and $h_+$  are given by $h_-(\vec{n}) = \mathrm{min}\{(\vec{n},\vec{X}): \vec{X} \in \mathcal{B}\}$ and
$h_+(\vec{n}) = \mathrm{max}\{(\vec{n},\vec{X}): \vec{X} \in \mathcal{B}\}$.
These quantities are Lipschitz-continuous functions on $\mathbb{S}^2$ (see e.g. Theorem 2.7 of \cite{HW}).  Note also that the quantity
$\int_{\mathcal{B}_{h,\vec{n}}}\!(\vec{n},\vec{X})d^3\!X$ appearing in (\ref{potential1}) is zero for $h \leq h_-(\vec{n})$ and
$h \geq h_+(\vec{n})$ and positive otherwise and for fixed $\vec{n}$ reaches its maximum at $h = 0$, i.e. when the center of gravity lies on the water plane. The positivity of $\int_{\mathcal{B}_{h,\vec{n}}}\!(\vec{n},\vec{X})d^3\!X$ when
$h_- (\vec{n}) < h < h_+(\vec{n})$ corresponds to the intuitively obvious fact that the center of buoyancy, namely the point $|\mathcal{B}_{h,\vec{n}}|^{- 1}\int_{\mathcal{B}_{h,\vec{n}}}\!\vec{X}\,d^3\!X$, lies below the center of gravity, namely the origin.\\ \\
We now turn to the issue of regularity. \\ \\
{\bf{Lemma 1}}: The functions on $\mathbb{R}^1 \times \mathbb{S}^2$ given by
$A = |\mathcal{B}_{h,\vec{n}}|$ and $B = \int_{\mathcal{B}_{h,\vec{n}}}\!(\vec{n},\vec{X})d^3\!X$ are $C^1$ for
$h_-(\vec{n}) < h < h_+(\vec{n})$, smooth (in fact: constant) for $h < h_-(\vec{n})$ as well as $h > h_+(\vec{n})$ and $C^{0,1}$ globally. The function
$C = h |\mathcal{B}_{h,\vec{n}}| - \int_{\mathcal{B}_{h,\vec{n}}}\!(\vec{n},\vec{X})d^3\!X$ is $C^2$ for $h_-(\vec{n}) < h < h_+(\vec{n})$  and $C^{1,1}$ globally.\\
We remark that when $\partial \mathcal{B}$ is smooth all these quantities are clearly smooth when $h_-(\vec{n}) < h < h_+(\vec{n})$. When in addition $\partial \mathcal{B}$ is smooth and has no flat portions, the first two are $C^{1,1}$ globally and the third one globally $C^{2,1}$ but not smoother.\\
{\bf{Proof of Lemma}}: The basic picture is that, when $h_-(\vec{n}) < h < h_+(\vec{n})$, by taking, for integrals of smooth functions over $\mathcal{B}_{h,\vec{n}}$, derivatives w.r. to $h$ or $\vec{n}$ one gets a volume term plus an integral of smooth functions over the planar figure
$\mathcal{S}_{h,\vec{n}} = \mathcal{B} \cap \{\vec{X}| h = (\vec{n},X) \}$. This is perhaps well known within the field of convex bodies, but we give a bare hands proof in the Appendix. Furthermore the surface terms are by inspection continuous and bounded near $h = h_+ (\vec{n})$ and  $h = h_- (\vec{n})$. Thus $A$ and $B$ are $C^1$ away from $h = h_+ (\vec{n})$ and  $h = h_- (\vec{n})$ with bounded derivatives, and $h = h_+ (\vec{n})$ and  $h = h_- (\vec{n})$ are Lipschitz. Consequently $A$ and $B$ are $C^{0,1}$. Finally the first derivatives of $C$ have no surface term. This ends the proof.\\ \\
We  collect the fundamental properties of the potential in\\ \\
{\bf{Theorem 1}}: The function $\mathcal{V}_s(h,\vec{n})$ is globally $C^{1,1}$. It is $C^2$ for  $h_-(\vec{n}) < h < h_+(\vec{n})$ and equal to $g\,h$ for $h > h_+(\vec{n}) > 0$ and equal to $g\,h(1 - \frac{1}{s})$ for $h < h_-(\vec{n}) < 0$. Furthermore $\mathcal{V}_s(h,\vec{n})$ is bounded from below for $s \leq 1$, in particular $\mathcal{V}_s(h,\vec{n}) \geq 0$ for $s \leq 1$ and  $\mathcal{V}_s(h,\vec{n}) > 0$ for $s < 1$.\\
{\bf{Proof}}: There only remains the statement on the lower bound. Clearly $\mathcal{V}_s$ is bounded below and achieves a global minimum which for $s<1$  has to occur at some value of $(h,\vec{n})$ for which $\partial_h \mathcal{V}_s$ is zero, in particular where $h_-(\vec{n}) < h < h_+(\vec{n})$ and where, by the second equation in (\ref{potential1}),
\begin{equation}
\partial_h \mathcal{V}_s = 1 - \frac{|\mathcal{B}_{h,\vec{n}}|}{s |\mathcal{B}|} = 0\,.
\end{equation}
Inserting this back into (\ref{potential1}) and recalling the positivity of $\int_{\mathcal{B}_{h,\vec{n}}}\!(\vec{n},\vec{X})d^3\!X$ when
$h_- (\vec{n}) < h < h_+(\vec{n})$ ends the proof for $s <1$. The case $s=1$ is similar. Here the minimum occurs for $h \leq h_-(\vec{n})$ and is zero. \\ \\
We now return to the equations of motion. These can be written as
\begin{equation}\label{h1}
\frac{d^2 h}{d t^2} = - \,\partial_h \mathcal{V}_s(h,\vec{n})
\end{equation}
and
\begin{equation}\label{Pi1}
\frac{d \vec{\Pi}}{dt}+ \vec{\Omega}  \times  \vec{\Pi} =  \vec{n} \times \partial_{\vec{n}} \mathcal{V}_s(h,\vec{n})
\end{equation}
together with
\begin{equation}\label{n1}
\frac{d \vec{n}}{d t} + \vec{\Omega} \times \vec{n} = 0\,.
\end{equation}
Remark: These equations do not form a Hamiltonian system on a symplectic space.
But their specific form expresses the fact that they  are Hamiltonian in the sense of a Poisson structure, namely the 'heavy top Poisson structure' (see \cite{MR}) with 2 symplectic dimensions added corresponding to the 'vertical' degree of freedom. We will however not pursue this viewpoint further.\\ \\
Clearly, when the body is up in the air, its vertical motion is constant acceleration pointing down. When the body is completely immersed, its vertical motion is acceleration pointing up when $s < 1$, pointing down when $s > 1$ and zero when $s = 1$. And, as is well known, a homogenous field exerts no torque on a rigid body in its rest frame. Thus, in these regions, the rotational motion is that of a free top. The interesting question for the dynamics is the interaction between the vertical and the rotational degrees of freedom and how this depends on the shape of the body (except for a spherical body, where there is no interaction whatsoever since $\mathcal{V}_s$ and $h_s$ are  independent of $\vec{n}$ in this case).   \\ \\
A consequence of the above findings is the\\ \\
{\bf{Theorem 2}}: Let $\mathcal{B}$ be convex.
\begin{itemize}
\item[(i)] There exists a unique solution of the system (\ref{h1}, \ref{Pi1}, \ref{n}) with initial values $(h(0), \dot{h}(0), \vec{\Pi}(0), \vec{n}(0))$.
\item[(ii)] When $s < 1$, the system stays in a bounded region of phase space.
\item[(iii)] When $s \geq 1$, solutions are still global in time and, for $s > 1$, have $\lim_{|t| \rightarrow \infty} h(t) = - \infty$.
\item[(iv)] When $s = 1$, either $h = \mathrm{const}$ or $\lim_{|t| \rightarrow \infty} h(t) = - \infty$.
\end{itemize}
{\bf{Proof}}: Local existence is standard (see e.g. \cite{HS}). The statement (ii) follows from the fact that the energy is the sum of three non-negative terms all of which blow up respectively in the unbounded variables $(\dot{h},\Pi, h)$. Statements (iii) immediately follows from $\ddot{h} \leq - g(1 - 1/s)$. Suppose $s = 1$: it is clear that $h$ has to ultimately enter the region $h \leq h_-(\vec{n})$ at $t = t_0$ or some $\vec{n}_0$. If $\dot{h}(t_0) < 0$, it will sink indefinitely. If $\dot{h}(t_0) = 0$, $\vec{n}_0$ can still perform rotational motions of a free top. Either the body as a result reenters the region $h > h_-(\vec{n})$, in which case it receives its final downward kick - or else $h$ stays constant and, by uniqueness, was also constant in the past. This proves (iv).   Global existence for $s \geq 1$ follows from
\begin{equation}
\dot{h}^2 + (I^{-1}\vec{\Pi},\vec{\Pi}) = 2 (E - \mathcal{V}_s(h,\vec{n}) ) \leq C_1 + C_2 |h|\,,
\end{equation}
where we have used that $|\mathcal{B}_{h,\vec{n}}|$ and $\int_{\mathcal{B}_{h,\vec{n}}}\!(\vec{n},\vec{X})d^3\!X$ are both bounded. Hence $|h|$ can at most grow quadratically in time and $|\vec{\Pi}|$ can at most grow linearly. \\ \\

\section{Equilibria and their stability}
We look at equilibria, namely the solutions given by
\begin{equation}
\frac{d h}{d t} = 0 =  \frac{d^2h}{d t^2} = 0\,,\,\,\,\frac{d \vec{n}}{d t} = 0\,,\,\,\, \vec{\Pi} = 0\,,\,\,\,\frac{d\vec{\Pi}}{d t} = 0
\end{equation}
Clearly $(h, \dot{h} = 0, \vec{\Pi} = 0, \vec{n})$ is an equilibrium iff
\begin{equation}\label{equi}
|\mathcal{B}_{\vec{n},h}| = s |\mathcal{B}|\,,\hspace{0.5cm}\vec{n}\times\!\int_{\mathcal{B}_{h,\vec{n}}}\!\!\!\!\!\!\! \vec{X}\, d^3X = 0\,.
\end{equation}
Those are exactly the conditions first formulated by Archimedes. Recall that we assume $\int_\mathcal{B}\vec{X}\,\, d^3X = 0$, so that the 2nd equation is the condition that   the line connecting the center of buoyancy with the centroid of the body is normal to the waterline. Note when the center of mass is submerged this does not automatically imply that this normal intersects the waterline inside the body. The first condition is sometimes in the literature called 'floating condition', the second 'equilibrium condition'. These conditions are equivalent respectively to $\partial_h \mathcal{V} = 0$ and
$\nabla_{n^A} \mathcal{V} = 0$, where $\nabla_{n^A}$ is the derivative tangential to $\mathbb{S}^2$, viewed as the submanifold $\{\vec{n} \in \mathbb{R}^3_\mathcal{B}|(\vec{n},\vec{n}) = 1\}$. So
$\nabla_{\vec{n}^A}$ acting on scalars is simply $H_A{}^B \partial_{n^B}$, where $H_A{}^B = \delta_A{}^B - n_A n^B$. Note the identity
\begin{equation}
\nabla_{n^A} (\vec{n}, \vec{X}) = H_{AB} X^B\,.
\end{equation}

In order to seek equilibria, we have to take $s < 1$ and solve the
floating condition in (\ref{equi}). Due to the monotonicity of $|\mathcal{B}_{h,\vec{n}}|$ for fixed $\vec{n}$ in $h$ for $h_-(\vec{n}) < h < h_+(\vec{n})$, there is a unique solution $h_s(\vec{n})$, which is $C^1$ in $\vec{n}$. In fact we find by implicit differentiation that
\begin{equation}\label{dnh}
\nabla_{n^A} \,h_s(\vec{n}) = |\mathcal{S}_{h_s(\vec{n}),\vec{n}}|^{- 1}\int_{\mathcal{S}_{h_s(\vec{n}),\vec{n}}}\!\!\!\!\!\!\! H_{AB} X^B \,d^2S (X)\,,
\end{equation}
where $\mathcal{S}_{h,\vec{n}} = \mathcal{B} \cap \{\vec{X}| h = (\vec{n},X) \}$. Now consider the function
\begin{equation}
\mathcal{U}_s(\vec{n}) = \mathcal{V}_s(h_s(\vec{n}),\vec{n}) = \frac{g}{s |\mathcal{B}|} \int_{\mathcal{B}_{h_s(\vec{n}),\vec{n}}}\!\!\!\!\!\!\!\!\!\!\!(\vec{n},\vec{X})\,  d^3\!X  \,,
\end{equation}
which by the above is $C^2$ (as opposed to $\int_{\mathcal{B}_{h,\vec{n}}}(\vec{n},\vec{X})\,  d^3\!X$, which is only $C^1$ in general). Let us remark that the closely result that the 'buoyancy surface' given by
\begin{equation}\nonumber
\frac{1}{s |\mathcal{B}|} \int_{\mathcal{B}_{h_s(\vec{n}),\vec{n}}}\!\!\!\!\!\!\!\!\!\!\!\vec{X}\,  d^3\!X
\end{equation}
is $C^1$ has been proved, using more sophisticated machinery, in Theorem 1.2. in \cite{HSW}.\\ \\
Coming back to $\mathcal{U}_s(\vec{n})$, it now follows that  equilibria are pairs $(h_s(\vec{n}_0), \vec{n}_0)$ with $\vec{n}_0$ a critical point of a $C^2$ - function on the compact manifold $\mathbb{S}^2$, which leads to the\\ \\
{\bf{Theorem 3}}: A convex body with $0 < s < 1$ has at least 2 equilibria. In fact, when equilibria are isolated, there holds the Poincar\'e-Hopf theorem, namely (see \cite{M})
\begin{equation}\label{Hopf}
N(\mathrm{max}) + N(\mathrm{min}) = N(\mathrm{saddle}) + 2\,.
\end{equation}
\\
We note at this point that there is a well known relation between equilibrium configurations and inverted ('capsized') configurations, due to the two identities:\\ \\
{\bf{Lemma 2}}: (i)\,$h_s(\vec{n}) = - h_{1 - s} (- \vec{n})$, \,\,
(ii) $\mathcal{U}_s(\vec{n}) = \frac{1 - s}{s} \,\mathcal{U}_{1 - s}
(- \vec{n})$.\\ \\
The identity (i) follows from
\begin{equation}
|\mathcal{B}_{h, \vec{n}}| + |\mathcal{B}_{- h, - \vec{n}}| = |\mathcal{B}|
\end{equation}
and then (ii) follows from
\begin{equation}
\int_{\mathcal{B}_{h,\vec{n}}}\!\!\!\!\!\!\vec{X} \,d^3X + \int_{\mathcal{B}_{- h,- \vec{n}}}\!\!\!\!\!\!\!\!\!\!\!\vec{X}\, d^3X = \vec{0}\,.
\end{equation}
It follows from (ii), that $\mathcal{U}_{\frac{1}{2}} (\vec{n}) = \mathcal{U}_{\frac{1}{2}} (- \vec{n})$ and from (i) that, for bodies with central symmetry, $h_{\frac{1}{2}} (\vec{n}) = 0$, i.e. centrally symmetric bodies with relative density 1/2 float with their centre of gravity at sea level.\\ \\
We remark that in the case of equilibria of convex, homogenous bodies with smooth boundary resting on a plane surface - which very heuristically can be regarded as the limit $s$ going to zero of the present situation - there holds the same formula (\ref{Hopf}), and here the question has been asked by Arnol'd, and recently answered affirmatively in \cite{VD}, if there exist bodies for which $N(\mathrm{max}) = N(\mathrm{min}) = 1$. The same question can of course be asked here: is there a body floating just in 2 configurations? For $\mathcal{B}$ a triaxial ellipsoid treated in Section 7, there holds $N(\mathrm{max}) = N(\mathrm{min}) = N (\mathrm{saddle}) = 2$.\\

Next note that, for $\mathcal{B}$ a ball, $\mathcal{U}_s(\vec{n})$ is constant: a sphere can float in every direction. A famous question by Ulam
\cite{U} has been if there are bodies other than the sphere for which this true (see \cite{R1} for a review). In fact, for $s = \frac{1}{2}$, such bodies have recently been constructed in \cite{V} and in \cite{R} for general dimensions.\\ \\
We now come to the issue of stability of equilibria. An isolated minimum of $\mathcal{V}_s(h,\vec{n})$ or $\mathcal{U}_s(\vec{n})$ is clearly stable, an isolated maximum unstable. For more information we have to compute the Hessian of the potential. First
\begin{equation}\label{heave}
\partial_h^2 \mathcal{V}_s = - \frac{g}{s |\mathcal{B}|}\, \partial_h |\mathcal{V}_{h,\vec{n}}| = \frac{g}{s |\mathcal{B}|}\, |\mathcal{S}_{h,\vec{n}}| > 0\,,
\end{equation}
where $\mathcal{S}_{h,\vec{n}}$ has been defined after (\ref{dnh}).
Physically, the r.h. side of (\ref{heave}) is $(\mathrm{frequency})^2$ of a ship's vertical oscillations around equilibrium. In particular ships are always stable against vertical perturbations. Furthermore
\begin{multline}\label{minusplus}
\nabla_{n^A} \nabla_{n^B} \mathcal{V}_s =
\frac{g}{s |\mathcal{B}|}\, \nabla_{n^A}\!\! \int_{\mathcal{B}_{h,\vec{n}}}\!\! H_{BC} X^C d^3X := H_A{}^{A'} H_B{}^{B'} \partial_{n^{A'}}
\nabla_{n^{B'}} \mathcal{V}_s = \\
= \frac{g}{s |\mathcal{B}|}\,\left(- H_{AB}\int_{\mathcal{B}_{h,\vec{n}}}\!\!\!(\vec{n},\vec{X})\, d^3 X +
\int_{\mathcal{S}_{h,\vec{n}}}\!\!\!\!\!\! H_{AD} X^D H_{BC} X^C \,d^2S(X) \right)\,.
\end{multline}
Thus for positive definiteness the positive surface term in (\ref{minusplus}) has to dominate the negative volume term. If that is the case we have the following\\ \\
{\bf{Theorem 4}}: Let the convex body $\mathcal{B}$ have $0 < s < 1$ and  $(h_s(\vec{n}_0), \vec{n}_0)$ be an equilibrium configuration for which the quadratic form in (\ref{minusplus}) is positive definite. Then this solution is (Liapunoff) stable, i.e. the time evolved solutions with initial data sufficiently near to
$(h(0) = h_s(\vec{n}_0), \dot{h}(0) = 0, \vec{\Pi}(0) = 0, \vec{n}(0) =
\vec{n}_0)$ remain close to the given one.\\
{\bf{Proof}}: The potential $\mathcal{V}_s(h,\vec{n})$, whence $\mathcal{H}$ is $C^2$ and has positive definite Hessian near the equilibrium configuration. Then the Taylor theorem applied to $\mathcal{H}$ and energy conservation say that $\mathcal{H}$ is a Liapunoff function, which (see e.g. \cite{HS}) implies stability.\\ \\
In order to make contact with the standard literature where one is interested in stability against specific kinds of perturbations we make the following observations. If $\vec{\delta}_0$ is a vector on $\mathbb{S}^2$ at the equilibrium point $\vec{n}_0$, we have
\begin{equation}\label{quadratic}
\nabla_{\vec{\delta}_0}\nabla_{\vec{\delta}_0}\mathcal{V}_s = \frac{g}{s |\mathcal{B}|}\,\left(-(\vec{\delta}_0,\vec{\delta}_0) \int_{\mathcal{B}_{h_s(\vec{n}_0),\vec{n}_0}}\!\!\!\!\!\!(\vec{\delta}_0,\vec{X})\, d^3 X +
\int_{\mathcal{S}_{h_s(\vec{n}_0),\vec{n}_0}}\!\!\!\!\!\! (\vec{\delta}_0,\vec{X})^2 \,d^2S(X)\right)\,.
\end{equation}
Let us suppose $(\vec{\delta}_0,\vec{\delta}_0) = 1$ and write, as is customary, $\vec{\delta}_0$ as
\begin{equation}
\vec{\delta}_0 = \vec{n}_0 \times \vec{t}
\end{equation}
with $\vec{t} \perp \vec{n}_0$ and $(\vec{t},\vec{t}) = 1$ so the vector $\vec{t}$ points along the axis of the infinitesimal rotation afforded by $\vec{\delta}_0$, yielding
\begin{equation}\label{surfaceinertia}
\nabla_{\vec{\delta}_0}\nabla_{\vec{\delta}_0} \mathcal{V}_s = \frac{g}{s |\mathcal{B}|}\,\left(- \int_{\mathcal{B}_{h_s(\vec{n}_0),\vec{n}_0}}\!\!\!\!\!\!(\vec{n}_0,\vec{X})\, d^3 X + \int_{\mathcal{S}_{h_s(\vec{n}_0),\vec{n}_0}}\!\!\!\!\!\!
\theta_{AB}\, t^A t^B d^2S(X)\right)\,,
\end{equation}
where $\theta_{AB}$ is given by
\begin{equation}
\theta_{AB} = \int_{\mathcal{S}_{h_s(\vec{n}_0),\vec{n}_0}}\!\!\!\!\!\! [\delta_{AB}\,(\vec{X},\vec{X}) - X_A X_B] d^2S(X)\,.
\end{equation}
So $\theta_{AB}$ is the  inertia tensor of the 2-dimensional figure in which the body intersects the water plane centered at the point at which the line through the centroid of $\mathcal{B}$ with tangent $n_0^A$ intersects the water plane. Let $\theta_1 > \theta_2$ be the eigenvalues of $\theta_{AB}$ with eigenvectors $t_1$ and $t_2$. Then the rotations associated with these axes are respectively called 'pitching' and 'rolling' in naval architecture.\\
The metacentre $\vec{M}(t)$ is defined by
\begin{equation}\label{meta}
\vec{M}(t) = \int_{\mathcal{B}_{h_s(\vec{\delta}_0),\vec{\delta}_0}}\!\!\!\!\!\!\!\!\vec{X}\, d^3 X - \vec{n}_0 \int_{\mathcal{S}_{h_s(\vec{n}_0),\vec{n}_0}}\!\!\!\!\!\!\!\! \theta_{AB} \,t^A t^B\,d^2S\,,
\end{equation}
and then stability under $\vec{\delta}_0 = \vec{n}_0 \times \vec{t}$ is given by the condition that the metacentre is above the centre of gravity, i.e. that $(\vec{n}_0, \vec{M}(t))$ be negative.\\ \\
Actually the metacentre defined in the literature is not based on the Hessian of $\mathcal{V}_s(h,\vec{n})$, but coincides with that based on the Hessian
of the potential $\mathcal{U}_s(\vec{n})$. Namely then the integrand of the surface term in (\ref{surfaceinertia}) becomes the inertia tensor w.r. to the centroid of the figure $\mathcal{S}_{h_s(\vec{n}_0),\vec{n}_0}$, so the associated metacentre lies in general below the one we have defined. In particular the condition used in our proof of Liapunoff stability is weaker than - i.e. implied by - that usually employed.
\section{The triaxial ellipsoid}
Here $\partial \mathcal{B}$ is given by the equation
\begin{equation}
q_{AB} X^A X^B = \frac{(X^1)^2}{a^2} + \frac{(X^2)^2}{b^2} +\frac{(X^3)^2}{c^2} = 1\,,
\end{equation}
where $a,b,c$ are positive numbers.  Let $\psi$ be the affine map from $\mathcal{B}' = B_1 (0)$, the unit-ball centered at the origin, to $\mathcal{B}$, with $\psi$ given by
\begin{equation}
X = \psi(Y) = \left(a\, Y^1, b\, Y^2, c\, Y^3\right)\,.
\end{equation}
Let $Q^{AB}$ be the inverse metric to $q_{AB}$ so that
\begin{equation}
Q^{AB}n_A n_B = a ^2 n_1^2 + b^2 n_2^2 + c^2 n_3^2\,.
\end{equation}
Then
\begin{equation}
\mathcal{B}_{h,\vec{n}} = \psi (\mathcal{B}'_{h',\vec{n}'}) \,\,\mathrm{with}\,\, h' = \frac{h}{(Q^{AB}n_A n_B)^\frac{1}{2}}\,,\,\,\,
n'_A = \frac{(a\, n_1, b\, n_2, c\, n_3)}{(q^{CD}n_C n_D)^\frac{1}{2}}
\end{equation}
and
\begin{equation}
\mathrm{det}(\psi) = (\mathrm{det}(q_{AB}))^{- \frac{1}{2}} = abc\,.
\end{equation}
Note that $\vec{n}'$ has again unit Euclidean norm. Thus
\begin{equation}
|\mathcal{B}_{h,\vec{n}}| = abc \,|\mathcal{B}'_{h',\vec{n}'}|
\end{equation}
and, since $h_-'(\vec{n}) = - 1$ and $h_+'(\vec{n}) = 1$,
\begin{equation}
h_-(\vec{n}) = - (Q^{AB}n_A n_B)^\frac{1}{2}\,,\hspace{0.5cm}h_+(\vec{n}) =  (Q^{AB}n_A n_B)^\frac{1}{2}
\end{equation}
and
\begin{equation}
\int_{\mathcal{B}_{h,\vec{n}}}\!\!(\vec{n},\vec{X})\, d^3X = abc \,
(a ^2 n_1^2 + b^2 n_2^2 + c^2 n_3^2)^\frac{1}{2}\int_{\mathcal{B}'_{h',\vec{n}'}}\!\!(\vec{n}',\vec{Y})\, d^3 Y\,.
\end{equation}
Next note that by spherical symmetry of $\mathcal{B}'$ both $|\mathcal{B}'_{h',\vec{n}'}|$ and $\int_{\mathcal{B}'_{h',\vec{n}'}}\!\!(\vec{n}',\vec{Y})\, d^3 Y$ are independent of $\vec{n}'$ and thus for $|h'| \leq 1$ there holds
\begin{equation}
|\mathcal{B}'_{h',\vec{n}'}| = 2 \pi \int_{h'}^1\!\! dZ \int_0^{\sqrt{1 - Z^2}}\!\!\!\! \rho\, d\rho = \frac{\pi}{3}\,(h'^3 - 3 h' + 2) = \frac{\pi}{3}(h'-1)^2(h'+2)
\end{equation}
and
\begin{equation}
\int_{\mathcal{B}'_{h',\vec{n}'}}\!\!(\vec{n}',\vec{Y})\, d^3 Y = 2 \pi \int_{h'}^1\!\! dZ\,Z \int_0^{\sqrt{1 - Z^2}}\!\!\!\! \rho\, d\rho = \frac{\pi}{4}\,(h'^2 - 1)^2\,.
\end{equation}
Thus, setting $\sigma =  (a ^2 n_1^2 + b^2 n_2^2 + c^2 n_3^2)^\frac{1}{2}$, we have for $|h| \leq \sigma$ that
\begin{multline}
\mathcal{V}_s (h, \vec{n}) = g \left[h - \frac{h}{4 s}\left(\left(\frac{h}{\sigma}\right)^3 - 3\, \frac{h}{\sigma} + 2\right) + \frac{3\, \sigma}{16 \,s} \left(\left(\frac{h}{\sigma}\right)^2 - 1\right)^2\right]=\\
= g \,\sigma \left[\frac{h}{\sigma} - \frac{1}{16 s}\left(\frac{h}{\sigma} - 1\right)^3\left(\frac{h}{\sigma} + 3\right)\right]
\end{multline}
Furthermore
\begin{equation}
\mathcal{V}_s (h, \vec{n}) = g\,h\,\,\,\,\mathrm{for}\,\,h > \sigma
\end{equation}
and
\begin{equation}
\mathcal{V}_s (h, \vec{n}) = - g\left(1 - \frac{1}{s}\right) h\,\,\,\,\mathrm{for}\,\,h <  - \sigma\,.
\end{equation}
Clearly $\mathcal{V}_s (h, \vec{n})$ is globally $C^{2,1}$ in $(h,\vec{n})$ and smooth for $|h| < \sigma$ (see the remark in the statement of Lemma 1).
To complete the model we have to compute $I_{AB}$. By similar methods we find that
\begin{equation}
I^{AB} = \frac{1}{|\mathcal{B}|} \int_\mathcal{B} [(\vec{X}, \vec{X}) \, \delta^{AB} - X^A X^B]\, d^3X = \frac{1}{5}\,(Q^{CD}\delta_{CD} \,\delta^{AB} - Q^{AB})\,.
\end{equation}
As for equilibria, we have
\begin{equation}
\mathcal{V}_{,h} = - \frac{g}{4\,s}\left[\left(\frac{h}{\sigma}\right)^3 - 3\, \frac{h}{\sigma} + 2(1 - 2 s) \right]\,
\end{equation}
for $|h| < \sigma$. Now the cubic $x^3 - 3 x + 2(1 - 2s)$ has 3 real roots, one of which lies in the interval $(- 1, 1)$: call this $x_s$. For example we have $x_0 = 1,\,x_{\frac{1}{2}} =0,\,x_1 = - 1$\footnote{More generally there holds $x_{1-s} = - x_s$ (see Lemma 2 and the remarks following it) and that $x_s$ is monotonically decreasing as a function of $s$.}. The potential $\mathcal{U}_s(\vec{n})$ takes the extremely simple form
\begin{equation}\label{s}
\mathcal{U}_s(\vec{n}) = \mathcal{V}_s(\sigma (\vec{n}) x_s, \vec{n}) = \frac{3\,g \,(x_s + 1)^2}{4 (x_s + 2)}\,\, \sigma (\vec{n})\,.
\end{equation}
Now there holds
\begin{equation}
\nabla_A \sigma = \frac{1}{\sigma} (\delta_{AB} - n_A n_B)\, Q^{BC} n_C = \frac{1}{\sigma} H_{AB}\, Q^{BC} n_C\,,
\end{equation}
so $n^A = n_0^A$ is a critical point of $\sigma$ iff it is an eigenvector of the linear map $Q^A{}_B$ (and thus also an axis of inertia). Let $n_0^A$ be such an eigenvector. Then
\begin{equation}
\nabla_A \nabla_B \sigma|_{n = n_0} = \frac{1}{\sigma}(H_{AC} H_{BD} - H_{AB}  n_C n_D)|_{n = n_0}Q^{CD} = \frac{Q_{AB} - \lambda(n_0)\, \delta_{AB}}{\sigma_0}\,,
\end{equation}
where $\lambda(n_0)$ is the eigenvalue of $Q$ associated with $n_0$. Recall $Q_{AB} = \mathrm{diag}(a^2,b^2,c^2)$ and $I_{AB} = (\frac{20 \pi abc}{3})^{- 1}\mathrm{diag}(b^2 + c^2, a ^2 + c^2, a^2 + b^2)$. Suppose $a > b > c$ and $\lambda(n_0) = b$. Then the first axis corresponds to the smallest eigenvalue of $I_{AB}$ and to the smaller eigenvalue of $\theta_{AB}$ whence to the rotation axis for rolling. Hence corresponding variations are along the third axis. And indeed
\begin{equation}
\nabla_A \nabla_B \sigma|_{n = n_0} = \frac{1}{b} \,\mathrm{diag}
(a^2 - b^2, 0, c^2 - b^2)\,,
\end{equation}
so we have instability against rolling and stability against pitching. And, of course, we have stability at the absolute minima $n_0^A = (0,0,\pm 1)$ and instability at the maxima $n_0^A = (\pm1, 0, 0)$.   \\ \\
Let us remark in passing that the expression (\ref{s}) has a finite nonzero limit given by $\frac{3 g}{8} \,\sigma(\vec{n})$ as $s$ tends to zero. On the other hand the potential for an ellipsoid lying on a plane surface is proportional to $r (\vec{n})$, the distance function relative the centre of mass. Clearly $\sigma(\vec{n}) = (Q^{AB} n_A n_B)^\frac{1}{2}$ and $r(\vec{n}) = (q_{AB} n^A n^B) ^{- \frac{1}{2}}$ are different, but at least have the same equilibria.  Is there a way to 'renormalize' the system so that the limit $s$ going to zero makes sense and where solutions somehow approach those of the anholonomic system describing an ellipsoid rolling on a plane?\\

Recall each equilbrium direction $\vec{n}_0$ is also parallel to an axis of inertia. Thus, looking at the equations (\ref{h},\ref{Pi},\ref{n}), for each of the corresponding equilibrium solutions, there is a 1-parameter class of solutions, which also rotate uniformly around $\vec{n}_0$ - and thus have nonzero angular momentum $(\vec{n}, \vec{\Pi})$ - i.e. of the form
\begin{equation}
\left(h(t) = \sigma (\vec{n}_0) x_s,\, \dot{h} = 0, \,\vec{\Pi}(t) \sim I \vec{n}_0 \sim \vec{n}_0, \,\vec{n} (t) = \vec{n}_0\right)\,.
\end{equation}
If $\vec{n}_0$ is along the shortest semiaxis these solutions will also be stable, since uniform rotation of a top along its longest and shortest axis of inertia is stable (see e.g. \cite{MR}) and there is no interaction between this rotation and the motion of $h$ and $\vec{n}$.\\ \\
There are more solutions one can find explicitly - up to quadrature -  namely those where the height degree of freedom $h$ and the directional degree of freedom $\vec{n}$ do not interact. These are first vertical motions with constant $\vec{n}_0$ relative to an equilibrium configuration $(h_s(\vec{n}_0)
, \vec{n}_0)$, e.g. vertical oscillations about a stable equilibrium. Second, in the case where $s = \frac{1}{2}$, there are spinning motions with constant $h = h_s(\vec{n}_0) = 0$ of $\vec{n}$ around any semiaxis. For example, let this rotation be w.r. to the first axis, i.e. set
\begin{equation}
\vec{n} = \cos\alpha \,\vec{f}_3 + \sin\alpha\, \vec{f}_2\,,\hspace{1cm}\vec{\Omega} = \dot{\alpha} \vec{f}_1\,,
\end{equation}
where $(\vec{f}_A, \vec{f}_B) = \delta_{AB}$, thus satisfying (\ref{n1}). Now (\ref{h1}) is valid for $h = 0$, and (\ref{Pi1}) becomes a 2nd order ODE for $\alpha$ with conserved energy given by
\begin{equation}
E = \frac{b^2 + c^2}{10}\,\dot{\alpha}^2 + \frac{3 g}{8}\,(b^2 \sin^2 \alpha + c^2 \cos^2 \alpha)^\frac{1}{2}\,.
\end{equation}
For example, when $E = \frac{3gb}{8}$, these are creeping orbits connecting the 2 equilibria at $\alpha = \frac{\pi}{2}$ and $\alpha = \frac{3 \pi}{2}$, which are unstable against rolling. We stop here leaving a fuller analysis to future work by us or others.\\ \\
A simpler system exhibiting some of the difficulties should be the 2 dimensional analogue of the present one of a floating ellipse - describing say an infinitely long floating log with elliptic cross section. Without giving details: the system in this case has just 2 degrees of freedom, namely the height $h$ and the directional degree $\vec{n} \in \mathbb{S}^1$ given by
$\vec{n} = (\cos \alpha, \sin \alpha)$, and the system is Hamiltonian in the proper sense. When $a, b$ are the semiaxes corresponding respectively to $\alpha = 0$ and $\alpha = \pi$, respectively to $\alpha = \frac{\pi}{2}$ and $\alpha = \frac{3 \pi}{2}$, the Hamiltonian is given - for simplicity we just state the case $s = \frac{1}{2}$ - by
\begin{equation}
\mathcal{H} = \frac{p_h^2}{2} + \frac{p_\alpha^2}{a b} + \mathcal{V}_\frac{1}{2}(h,\alpha)\,,
\end{equation}
where
\begin{equation}
\mathcal{V}_\frac{1}{2}(h,\alpha) = \frac{2 g\, h}{\pi}\left(\frac{h}{\sigma}\,\sqrt{1 - \frac{h^2}{\sigma^2}} + \arcsin \frac{h}{\sigma}\right) +
\frac{4 g\, \sigma}{3 \pi}\,\left(1 - \frac{h^2}{\sigma^2}\right)^\frac{3}{2}
\end{equation}
for $- \sigma \leq h \leq \sigma$ and where $\sigma(\alpha) = \sqrt{a ^2 \cos^2 \alpha + b^2 \sin^2 \alpha}$, and
\begin{equation}
\mathcal{V}_\frac{1}{2}(h,\alpha) = g |h|\hspace{0.6cm}\mathrm{for}\,\,|h| \geq \mu
\end{equation}
Again the potential is $C^{2,1}$. The equations of motion are given by
\begin{equation}
\frac{d h}{d t} = \frac{\partial \mathcal{H}}{\partial p_h}\,,\,\,\,\,\,\,\frac{d p_h}{d t} = - \frac{\partial \mathcal{H}}{\partial h}\,,\,\,\,\,\,\,\frac{d \alpha}{d t} = \frac{\partial \mathcal{H}}{\partial p_\alpha}\,,\,\,\,\,\,\,\frac{d p_\alpha}{d t} = - \frac{\partial \mathcal{H}}{\partial \alpha}\,.
\end{equation}
\section{Final remarks}
We have in this paper introduced and studied a conservative system of ordinary differential equations obeyed by a rigid body which is subject to the combined action of a constant gravitational field and the static buoyancy force coming from the Archimedes principle. This system is the time dependent version of the equilibrium theory of floating bodies studied by mathematicians, physicists and naval engineers over the centuries. Our main result has been the Liapunoff stability of those equilibria, which are minima of the potential $\mathcal{V}_s(h,\vec{n})$. Although we have no new results in the equilibrium theory itself, we hope that the potential $\mathcal{V}_s (h, \vec{n})$, which has been our main tool, will also be useful for open issues in the equilibrium theory. For the dynamics we believe we have only touched the surface of a rich structure. Further progress using analytical and numerical means should be possible. A potential challenge, at least on the numerical side, could be the fact that the potential is at most $C^{2,1}$ at points in phase space which correspond to the body touching the water line from above or below.\\ \\

\section{Appendix}
We consider integrals of the form
\begin{equation}\label{Ihn}
I_{h,\vec{n}} = \int_{\mathcal{B}_{h,\vec{n}}}\!\!\!\!f(\vec{X})\, d^3X\,,
\end{equation}
where $\mathcal{B}$ is a convex body, $h_-(\vec{n}) < h < h_+(\vec{n})$ and $f$ is a smooth function on $\mathcal{B}$. We want to show that $I$ has continuous derivatives w.r. to $(h,\vec{n})$, namely that
\begin{equation}
\partial_h I (h, \vec{n}) = -  \int_{\mathcal{S}_{h,\vec{n}}}\!\!\!\!f(\vec{X})\, d^2S
\end{equation}
and for some family $\vec{n}_\alpha \in \mathbb{S}^2$ with $|\alpha|$ small and $\vec{n}_0 = \vec{n}$,
\begin{equation}\label{other}
I' (h, \vec{n}_\alpha)|_{\alpha = 0} =  \int_{\mathcal{S}_{h,\vec{n}}} \!\!\!(\vec{n}_0{}', \vec{X}) f(\vec{X})\, d^2S\,,
\end{equation}
where a prime denotes derivative w.r. to $\alpha$. It suffices to consider integrals of the form (\ref{Ihn}) where  $\mathcal{B}_{h,\vec{n}}$ is replaced by the layer $\mathcal{B}_{h,\vec{n}} \setminus \mathcal{B}_{\bar{h},\vec{n}}$ for some fixed $\bar{h}$ with $h < \bar{h} < h_+(\vec{n})$. We call these integrals $J_{h,\vec{n}}$. Now assume without loss that $\vec{n} = (0,0,-1)$. Taking $\bar{h}$ sufficiently close to $h$, there is a vertical (i.e. parallel-to-$\vec{n}$-) line which intersects both  $\{(\vec{n},\vec{X}) = \bar{h}\}$ and $\{(\vec{n},\vec{X}) = h\}$ and all parallel planes in between. We then choose cylindrical coordinates $(\rho, \phi, Z)$ for this central line with $Z = - (\vec{n},\vec{X})$. Now any plane through this central line intersects the boundary of the convex body  $\mathcal{B}$ in a curve which is again convex and so between these planes $\partial \mathcal{B}$ can be written as $\rho = R(\phi,Z)$, where $R$ is strictly positive, continuous in $\phi$ (in fact: $C^{0,1}$ in $\phi$ since the boundary of a convex body is $C^{0,1}$, but we do not use this at this point) and convex whence $C^{0,1}$ in $Z$. There is thus a function $R(\phi,Z)$ such that
the domain of integration is given by $0 \leq \phi \leq 2 \pi,\,- \bar{h} \leq Z \leq h,\,0 \leq \rho \leq R(\phi,Z)$ and the integral given by
\begin{equation}
 = \int_{- \bar{h}}^{- h} \!\!\! dZ\,\int_0^{2 \pi}\!\!\! d\phi\, \int_0^{R(\phi,Z)}\!\!\!\!\!\!\rho\,f(\rho, \phi, Z)\, d\rho\,,
\end{equation}
from which the statement on $\partial_h$ follows. Proving Eq.(\ref{other}) is a little more tricky. Here the domain of integration is given by $0 \leq \phi \leq 2 \pi,\,- \bar{h} \leq Z\,,- (\vec{n}_\alpha,\vec{X})\leq - h\,,0 \leq \rho \leq R(\phi,Z)$. We first set $\rho = \bar{\rho} \,R(\phi, Z)$ and without loss choose $\vec{n}_\alpha = (0, - \sin \alpha, - \cos \alpha)$. The domain of integration is given by $0 \leq \phi \leq 2 \pi,\,- \bar{h} \leq Z\,,- (\vec{n}_\alpha,\vec{X}) = Z \cos\alpha + \bar{\rho} \,R(\phi, Z) \sin\phi \sin \alpha \leq - h\,,0 \leq \bar{\rho} \leq 1$. We now for small $|\alpha|$ seek a function $Z_\alpha(\phi, \bar{\rho}, h)$ with $Z_0 (\phi, \bar{\rho}, h) = - h$ such that
\begin{equation}\label{implicit}
Z_\alpha \cos\alpha + \bar{\rho} \,R(\phi, Z_\alpha) \sin\phi \sin \alpha = - h\,.
\end{equation}
To solve (\ref{implicit}) the implicit function theorem can not be used, since $R$ is only Lipschitz in $Z$. But that is still good enough in this case, as the Lemma below shows. It also shows that $Z_\alpha$ is differentiable at $\alpha = 0$, and so $Z_0'(\phi, \bar{\rho}, h) = -
\bar{\rho} \,R(\phi, - h) \sin\phi$. The integral is thus given by
\begin{equation}\label{diff}
J_{h,\vec{n}_\alpha} = \int_0^{2 \pi}\!\! d\phi \int_0^1 \!\!\bar{\rho}\, d \bar{\rho}
\int_{- \bar{h}}^{Z_\alpha(\phi,\bar{\rho},h)}\!\! R^2(\phi, Z)\, \bar{f}(\bar{\rho}, \phi, Z)\,dZ
\end{equation}
Now the statement follows by differentiating (\ref{diff}) w.r. to $\alpha$ under the integral sign and noting that
\begin{equation}
Z_0'(\phi, \bar{\rho}, h) = (\vec{n}_0{}', \vec{X})|_{\mathcal{S}_{h,\vec{n}}}\,.
\end{equation}
Finally the solution to (\ref{implicit}) can for small $|\alpha|$ be reduced to the following\\ \\
{\bf{Lemma}}: Consider the equation
\begin{equation}\label{implicit1}
Z = \alpha f(Z,\alpha) + g(\alpha)\,,
\end{equation}
where $f,g$ are smooth in $\alpha$ and $f$ is Lipschitz in $Z$ near $Z_0 = g(0)$ with Lipschitz constant $K$. Then (\ref{implicit1}) has a unique family
of solutions $Z_\alpha$ with $Z_0 = g(0)$. Furthermore $Z_0{}' = f(Z_0,0) + g'(0)$.\\
{\bf{Proof}}: Taking $0 < \alpha < \frac{1}{K}$ the r.h. side of (\ref{implicit1}) defines a contraction map for $Z$ near $Z_0$, so $Z_\alpha$ exists by virtue of the Banach fixed point theorem (see e.g. \cite{Sh}). Then, inserting $Z_\alpha$ back into (\ref{implicit1}) and taking the difference quotient near $\alpha = 0$, gives the last statement.\\ \\
{\bf{Acknowledgments}}: One of us (R.B.) thanks Helmuth Urbantke for discussions in the initial stage of this work. We furthermore thank an anonymous referee for pointing out the relevance of reference \cite{HSW}.

\end{document}